# Waves on the surface of the Orion Molecular Cloud

Olivier Berné[1,2], Núria Marcelino[1], José Cernicharo[1]

[1]*Centro de astrobiología, CSIC/INTA,Ctra de Torrejón a Ajalvir, km 4 28850 Torrejón de Ardoz, Madrid Spain*

[2]*Present address Leiden Observatory, Leiden University, P.O. Box 9513, NL- 2300 RA Leiden, The Netherlands*

**Massive stars influence their parental molecular cloud, and it has long been suspected that the development of hydrodynamical instabilities can compress or fragment the cloud[1,2]. Identifying such instabilities has proven difficult. It has been suggested that elongated structures (such as the 'Pillars of creation'[3]) and other shapes arise because of instabilities[4,5], but alternative explanations are available[6,7]. One key signature of an instability is wave-like structures in the gas, which have hitherto not been seen. Here we report the presence of 'waves' at the surface of the Orion cloud near where massive stars are forming. The waves appear to be a Kelvin-Helmholtz instability, arising during the expansion of the nebula, as gas heated and ionised by massive stars is blown over pre-existing molecular gas.**

Because it is the closest massive star forming region (d~414 pc[8]), the Orion nebula provides a unique opportunity to study at small spatial scales, combining different wavelengths, the feedback action of the HII region and plasma on the cloud[9,10]. Figure 1 presents the carbon monoxide (CO) millimetre map of the cloud, that has a spatial coverage and angular resolution enabling morphological comparison with high angular resolution mid-infrared and X-Ray images.

In the southern part of the nebula, the growing bubble of gas ionised by massive stars (HII region) has pushed the surrounding molecular gas towards the far and near side of the nebula (respectively red and blue in Fig. 1.b.). The X ray plasma bubble[12], fills the southern cavity of molecular gas (Fig. 1.b.). In the infrared, the cloud surfaces exhibit «ripples» in several regions. This is particularly obvious in the Southwest region, (Fig. 1.c.) where a series of 5 surprisingly regular wavelets (the Ripples hereafter) enshrouding an elongated molecular cloud, subject to a strong velocity gradient (7-9 km.s$^{-1}$.pc$^{-1}$, Fig. 1.c and 2.a), are seen.

A possibility would be that the CO structure results from the outflow of a young protostar, and the IR emission from the associated Herbig-Haro objects[13,14,15]. However, we do not find any evidence of an embedded protostar able to drive an outflow in this region, neither in the Spitzer IRAC/MIPS images nor in the SIMBAD database. Nor do we find evidence of wings in the line profiles of $^{12}$CO emission (Fig. 2a), which are characteristic of outflows[16]. Examining the Spitzer infrared spectrograph (IRS) archival spectrum of the second Ripple (Fig. 2c), we find no trace of highly ionised species ([NeIII], [FeII]), that would attest to the presence of a high velocity Herbig-Haro shock[13,14,15] (the [SIII] emission detected here is spatially associated to the HII region). The comparison between the spatial distribution of Polycyclic Aromatic Hydrocarbons (PAHs), $H_2$ 0-0 S(1) pure rotational line intensity, and $^{12}$CO (2-1) emissions in an East to West cut across the Ripples (marked 2 in Fig. 2b) shows a stratification with PAHs more to the East, followed by $H_2$ and then CO emission (Fig. 2d). This stratification is expected in UV driven Photo-Dissociation Regions (PDR) which lie at the boundary between HII regions formed by massive stars and molecular gas[17,18], but not in shocks. Furthermore, the $H_2$ S(1) intensity (~$7\times10^{-8}$ W.m$^{-2}$.sr$^{-1}$) is compatible with the predictions of PDR models[19] reproducing the conditions of the observed region: dense gas (above $10^4$ cm$^{-3}$) highly irradiated ($10^3$ times the interstellar standard radiation field). We therefore conclude that the observed periodic structure and emission are not the result of an outflow/Herbig-Haro shock, but are rather consistent with a succession of PDRs lying at the surface of an elongated, rippled, molecular cloud, shaped in a smooth process (Fig. 3).

The rippled molecular cloud is subject to an important velocity gradient (7-9 km.s$^{-1}$.pc$^{-1}$, Fig. 1 and 2). This gradient cannot be due to the simple expansion of the HII region (that is ~3 km.s$^{-1}$ for a spherical bubble of several pc, based on the difference between mean velocities of the blue and red parts of the southern cloud), so it must result from the acceleration of the cloud by the passage of a flow of diffuse gas. We therefore propose that the Ripples have been formed by the mechanical interaction of high velocity plasma/gas produced by massive stars with the dense molecular gas, which has provoked hydrodynamical instabilities. The simplest non-trivial hydrodynamical instabilities able to explain the observed structures are Rayleigh-Taylor (RT) or Kelvin-Helmholtz (KH) instabilities[22]. The RT instability occurs in an interface between fluids of different densities, while the KH instability needs an additional velocity difference between these fluids. In the present case, where density and velocity gradients are present, the KH instability will dominate (see sup. material). In addition to the hydrodynamical phenomenon, the effect of energetic radiation has to be taken into account. Far and extreme ultraviolet (respectively FUV and EUV) photons emitted by the massive stars will create a photo-ablation layer at the surface of the cloud that will insulate the molecular cloud from the sheering flow and prevent the instability from developing.

In order to verify that the development of a Kelvin-Helmholtz instability is possible, let us first consider a purely hydrodynamical case, where the insulating layer is assumed to be infinitely thin. Kelvin-Helmholtz instabilities having large spatial wavelengths will be removed by gravity[22] above a value $\lambda_{KH}$, which depends upon (see sup. material), the self gravity $g_c$ of the cloud, the density $n_f$ and velocity $v_f$ of the sheering gas, and the density of the sheered gas in the cloud $n_c$. The CO observations allow us to estimate the cloud column density, which, assuming a cylindrical geometry constrains the density $n_c$ and gravity $g_c$ (resp. $10^4$ cm$^{-3}$ and 3.5x10$^{-11}$ m.s$^{-2}$) so that $\lambda_{KH}$ can be estimated only from the parameters $n_f$ and $v_f$. Similarly, the growth rate $\omega_{KH}$ of a Kelvin-Helmholtz instability of known spatial wavelength $\lambda$ solely depends on $n_f$ and $v_f$[22] (sup. information). The observed value of $\lambda$=0.11 pc sets a minimum for $\lambda_{KH}$ and the observed width of the CO lines (~2 km.s$^{-1}$) and height of the ripples (unlikely to be more than about 0.1 pc) imply a maximum growth rate $\omega_{KH}$~2x10$^{-5}$ yrs$^{-1}$. The analysis of the evolution of $\lambda_{KH}$ and $\omega_{KH}$ using a grid of realistic values of $n_f$ and $v_f$, show that typical properties of either HII gas or X ray plasma result in values for $\lambda_{KH}$ and $\omega_{KH}$ that respect the observational constraints, and allow an instability to grow.

Now let us consider the effect of UV radiation, that will allow the propagation of the instability to the entire cloud only if the instability wavelength is larger than the thickness of the photo-ablation layer, consisting in the sum of a FUV dominated sub-layer and a EUV dominated sub-layer (Figure 2. in sup. information). The FUV dominated sub-layer thickness, estimated from the PAH emission in the IRAC images, is likely small (~0.01 pc). However, estimates show (sup. information) that in the presence of Theta 1 Ori C, the EUV dominated sub-layer is too large to allow the propagation of a KH instability to the cloud. Given that the growth time for the observed instability is estimated to be above $10^5$ years[22], it is well possible that it formed before the birth of Theta 1 Ori C. At that time, the Lyman continuum photon flux would have been significantly lower than today, and so were, as a consequence, the density and thickness of the EUV dominated layer. With a much smaller insulating layer and in the absence of the X-ray plasma (that can only be driven by Ori C), it would have been the expansion of dense HII gas produced by the population of older massive stars that resulted in the shearing of a pre-existing molecular clump, forming the KH instability at the position where we observe the Ripples. Because the surface of the cloud considered here is rippled and tilted, the later introduction of intense irradiation after the birth of Theta 1 Ori C would have resulted in an heterogenous rocket acceleration[23,24] of the clumps, possibly taking over the KH instability.

This selective photo-abrasion is now generating a train of molecular globules where the most Eastern part of the elongated cloud is already detached from the rest of the cloud (Fig. 2.b.). The masses of the globules, close to 1 $M_{Sun}$, imply that they can survive several $10^5$ years to photo-ablation[25] (see sup. informations). Such a scenario for the formation of the Ripples is supported 1) by the observation of bended outflows in Orion[26], attesting for the run-over of HII gas on pre-existing molecular gas hosting low mass star formation and 2) the periodicity in the distribution of molecular clumps in Orion's molecular filaments, that may result from similar Kelvin-Helmholtz induced fragmentation[5]. Additional observations of periodic structures in Orion or other massive star forming regions at high angular and spectral resolution (in particular at millimetre and radio wavelengths), could contribute to obtaining a clearer picture of the history of combined radiative/hydrodynamical feedback of massive stars on their parental cloud.

**Supplementary Information** accompanies the paper on **www.nature.com/nature**.

**Acknowledgements** The authors wish to thank J. Bally and M. Pound, for their comments which contributed to the improvement of the manuscript. The authors wish to aknowledge Bertrand Lefloc'h, Javier Goicoechea and Jesús Martín-Pintado for critical discussions. David Hochberg is aknowledged for his reading of the manuscript. O. B. acknowledges C. Joblin and Prof. S. B. at UofP for support.

**Author contribution** Olivier Berné conducted the scientific analysis and write-up of the paper. Nuria Marcelino conducted the CO observations and analysis. J. Cernicharo was at the origin of this project and its principal investigator.

**Author information** Reprints and permissions information is available at npg.nature.com/reprintsandpermissions. Correspondence and requests for materials should be addressed to O.B. (e-mail: berne@strw.leidenuniv.nl).

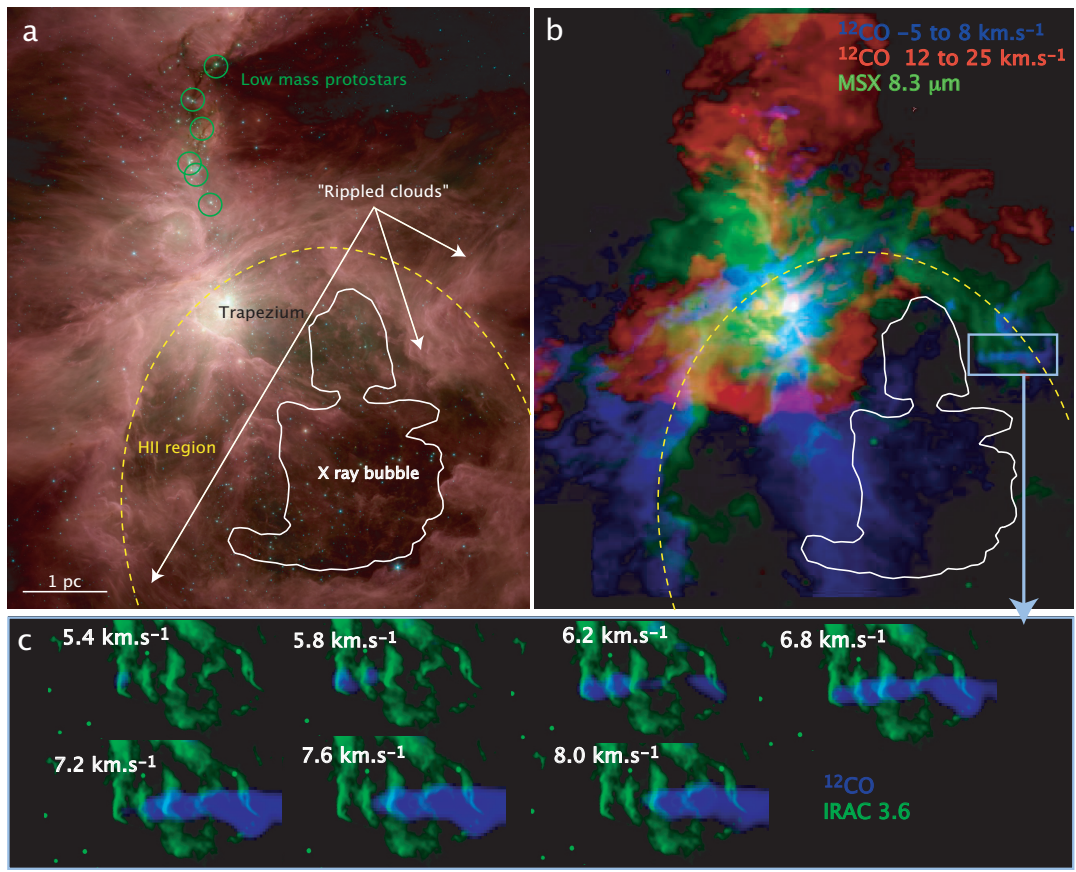

Fig. 1 **Multi-wavelength overview of the Orion nebula**. **a,** Spitzer mid-infrared image of the Orion nebula [NASA JPL/Caltech /T. Megeath]. Green circles indicate the positions of protostars. **b,** Overlay of the IRAM 30m $^{12}$CO J=2-1 maps integrated emission in velocity between -5 and 8 km.s$^{-1}$ (blue), and 12 and 25 km.s$^{-1}$ (red), tracing the cold (10-100 K) molecular gas, and 8 μm emission from MSX (green, 18'' resolution) tracing dust and warm (100-10$^3$ K) gas at the surface of clouds. The CO observations were performed using the HERA multibeam receiver[11], tuned at the frequencies of $^{12}$CO and $^{13}$CO J=2-1 (resp. 230.5 GHz and 220.5 GHz) in each polarisation. The spatial resolution of the maps is 12 " (0.025 pc) and the spectral resolution is 0.4 km.s$^{-1}$. **c**, A close-up view of the Ripples in $^{12}$CO (2-1) emission at different velocities overlaid on the Spitzer IRAC 3.6 μm map (attributed to PAH emission, green, 2.5" resolution). In panels **a** and **b**, the white solid contours show the delimitation of diffuse soft X-ray emission tracing the hot plasma detected by XMM[12], and the yellow dashed contour shows the delimitation of the HII region.

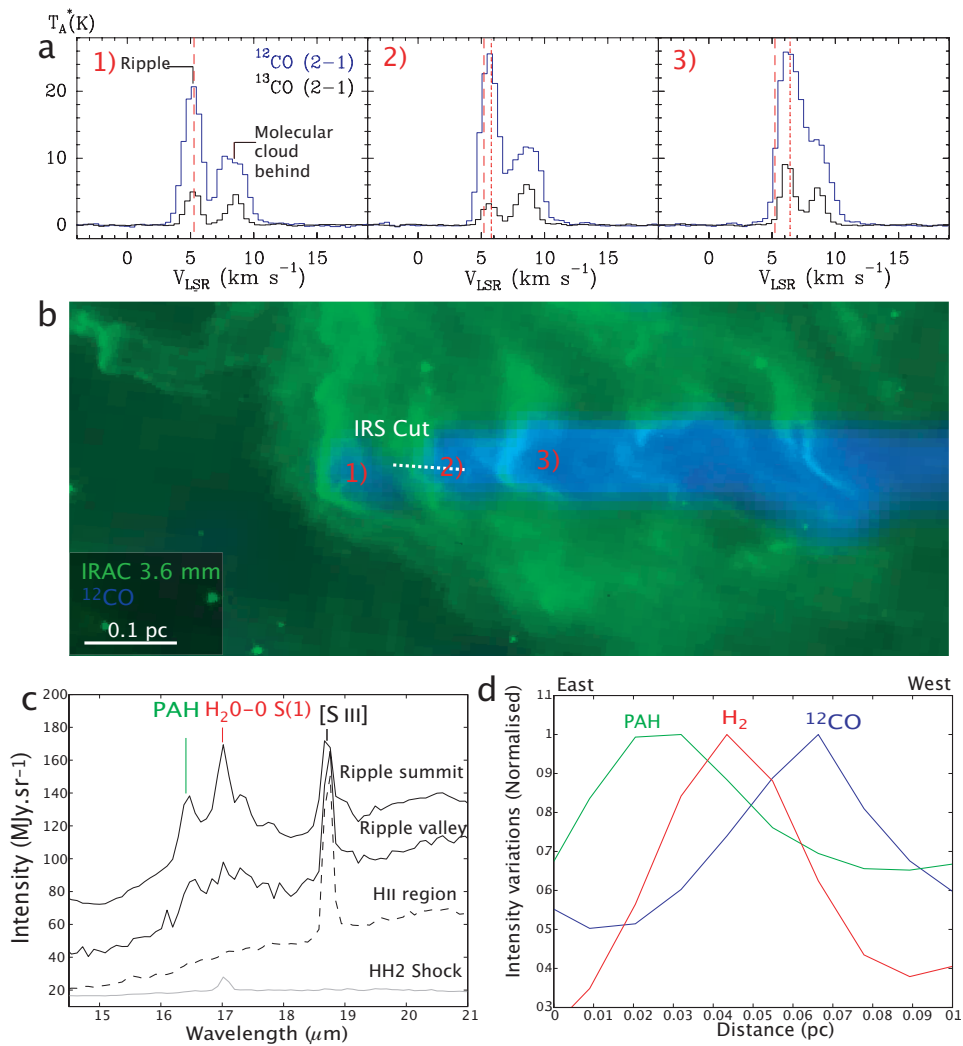

Fig. 2 **Infrared and millimetre observations of the Ripples**. **a,** IRAM-HERA spectra of the $^{12}$CO and $^{13}$CO (2-1) lines at the three positions indicated in the image. The red vertical dashed line indicates the velocity of Ripple 1) and dotted lines the velocity of Ripples 2) and 3). **b,** General view of the Ripples and surrounding environment, Spitzer-IRAC 3.6 µm image in green, $^{12}$CO integrated over all velocities overlaid in blue. **c,** Spitzer IRS-LL2 spectra of the summit and valley of Ripple 2 and surrounding HII region. The LL2 spectrum of the Herbig Haro 2 (HH2) shock is given for comparison. **d,** East to West cut across a Ripple showing the evolution of the line integrated emission of PAH, H$_2$ and $^{12}$CO presented in the spectra.

.

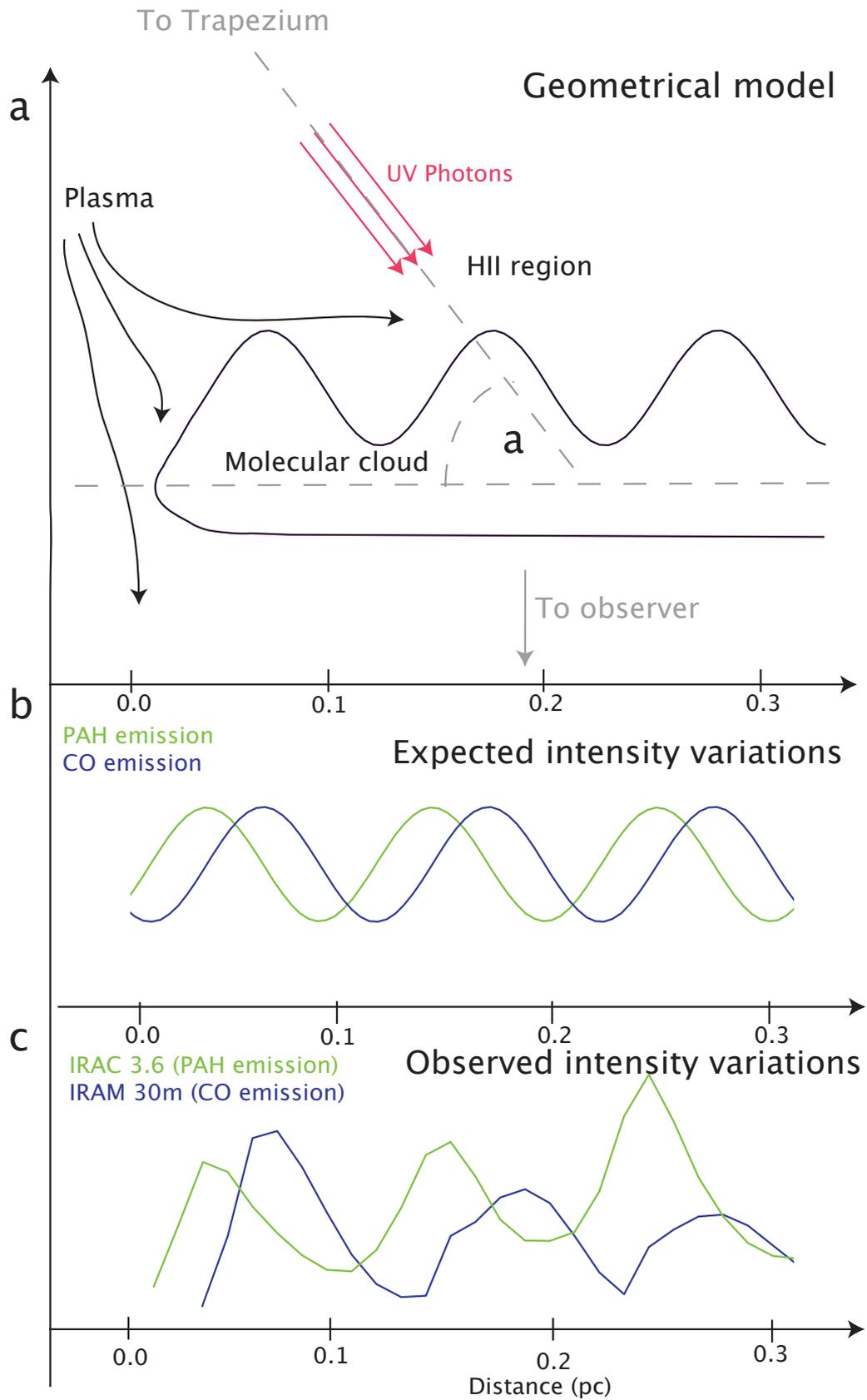

Fig. 3 **Geometrical model of the Ripples**. **a,** Schematic view of the geometric configuration. **b,** Expected PAH and $^{12}$CO emission and **c,** observed PAH and $^{12}$CO emission with IRAM and IRAC. We assume that energy collected in the IRAC 3.6 μm broad band filter is dominated by the 3.3 μm band of PAHs. This is likely the case as the 16.4 μm PAH band is strongly detected in spectroscopy. The displacement of peak emission between the two tracers can be well explained by a configuration in which the major axis of an elongated cloud forms an angle α of less than 90 degrees with the direction pointing to the trapezium stars: the energy per unit of surface emitted by PAH molecules at the surface of the cloud, which depends on the UV illumination, is stronger on the upward side of the ripple, and weaker on the downward shadowed part. On the other hand, the intensity of CO emission depends on the column density of gas in the line of sight, so it peaks at the summit of the Ripple. The observed CO intensity gives a column density of gas $N_{H_2} \sim 6 \times 10^{21}$ cm$^{-2}$ (see sup. material for details) corresponding to an optical depth[20] at 3.6 μm of ~$10^{-2}$. This allows one to see perfectly through the cloud in the mid-infrared. This geometry is compatible with the Ripples being in the blue-shifted part of the cloud (Fig. 1), in front of the HII region, and with the velocity gradient due to the passage of the flow being from East to West. This configuration could also be an explanation for the non detection of the plasma in X rays due to extinction at this position[12,21].